\documentclass{article}

\usepackage{latexsym}

\begin{document}

\newtheorem{theorem}{Theorem}
\newtheorem{proposition}{Proposition}
\newtheorem{remark}{Remark}
\newtheorem{corollary}{Corollary}
\newtheorem{lemma}{Lemma}
\newtheorem{observation}{Observation}
\newtheorem{definition}{Definition}

\newcommand{\qed}{\hfill$\Box$\medskip}

\title{An NL-Complete Puzzle} 
\author{Holger Petersen\\
Reinsburgstr. 75\\
70197 Stuttgart\\
Germany 
}

\maketitle

\begin{abstract}
We investigate the complexity of  a puzzle that turns out to be {NL}-complete.
\end{abstract}

\section{Introduction}

In this note we discuss one of  24 different mathematical
puzzles appearing under the title ``Kopfnuss'' (literally clout, but used here in the sense of 
a brain teaser) in the newspaper tz (see also \cite{BullsPress2015} for examples).
Among these are well-known classical puzzles like Sudoku and Kakuro. 
Some are tiling puzzles in the style of Tangram that do not offer
an obvious way of generalizing them to unbounded sizes. Others however can be 
turned into decision problems that can be investigated from a Complexity Theory perspective. 

David Eppstein \cite{Eppstein} states: 
\begin{quote}
If a game is in P, it becomes no fun once
you learn ``the trick'' to perfect play\ldots\\
To me, the best puzzles are NP-complete (although some good puzzles 
are in P, relying on gaps in human intuition rather than on computational 
complexity for their difficulty).
\end{quote}
Here we challenge these observations by presenting a real puzzles (in the sense that
it was invented to be played and in fact published in a newspaper) which turn out 
to be {NL}-complete when generalized to unbounded size.

\section{Result}

The problem is called ``The Way Through the Purgatory'' 
(``Der {W}eg durchs {F}egefeuer'', \cite{tz20100419,tz20101018,tz20130610})
and 
asks for a path along a list of $n$ numbers (in published instances of the puzzle
$n = 53$ and the list is represented as a spiral). The path starts at the first
number and continues from number $i$ at position $j$ to number $j-i$ or number
$j+i$, provided that the resulting values are in the range from  $1$ to $n+1$. If the result
is $n+1$, the puzzle is solved, since the new position is just after the list.
As an example consider the list:
$$\ell = (3, 2, 2, 1, 4, 2, 1, 2, 3)$$
A solution is $1, 4, 5, 9, 6, 8, 10$.

We consider here the decision problem ``The Way Through the Purgatory'', consisting of all lists
with numbers encoded in binary having a solution.

\begin{theorem}
The problem ``The Way Through the Purgatory'' is complete in {NL}.
\end{theorem}
{\bf Proof.} First we show membership in {NL}. A nondeterministic TM $M$
marks $(\log n)+1$ cells on its tape. Starting on the first number of the input, 
$M$ iteratively copies a number (if it fits into the marked space) 
to the work-tape and nondeterministically decides whether to move 
forward or backward. It then uses the stored value for counting 
the moves on the input. If the input head is about to leave the input at the left boundary, 
$M$ rejects. If it leaves the input at the right boundary, $M$ accepts
exactly if the count reaches zero. Notice that numbers exceeding $n$ need not be copied to the 
tape, since they cannot point to a position in the list.

For hardness we reduce the problem PATH \cite[p.~326 ff]{Sipser06} to the 
given problem, where 
$$\mbox{PATH} = \{ \langle G, s, t\rangle 
\mid \mbox{$G$ is a directed graph that has a path from $s$ to $t$}\}.$$ 
Let $\langle G, s, t\rangle$ be an instance of PATH
with $m$ vertices. We first reduce $\langle G, s, t\rangle$ to 
$\langle G', u_1, u_n\rangle$ by replacing each vertex $u$ of outdegree
$d > 2$ with a path consisting of $d-1$ vertices starting at $u$
and including $d-2$ new vertices, where from the last vertex two
target vertices are reachable and from the others one target is reachable.

As an example take edges $(u, v_1)$,  $(u, v_2)$, and $(u, v_3)$. We
introduce new vertex $u_1$, remove $(u, v_2)$ and $(u, v_3)$, and introduce
$(u, u_1)$, $(u_1, v_2)$, and $(u_1, v_3)$. Notice that the size of the 
resulting graph with $n$ vertices is linearly bounded and reachabilty is not affected by ths construction.

Now we reduce $\langle G', u_1, u_n\rangle$ to the purgatory problem. We
concatenate several lists of numbers. The first list contains a number per
vertex and has the form:
$$(4n+1, 4n+3, \ldots, 4n+2i-1, \ldots, 6n-1)$$
These numbers point to the central element of a list of three numbers
encoding at most two edges of $G'$.

The second list consists of $3n$ copies of the number $7n$.

Then we add $n-1$ lists of length $3$ each, where list $\ell_i$ takes one of
three possible forms depending on $u_i$:
\begin{description}
\item[\boldmath$u_i$ has outdegree 0:] $\ell_i = (7n, 7n, 7n)$
\item[\boldmath$u_i$ has outdegree 1, $G'$ contains edge $(u_i, u_j)$:]
    $\ell_i = (3i+4n-j-3, 1, 7n)$
\item[\boldmath$u_i$ has outdegree 2, $G'$ contains edges $(u_i, u_j), (u_i, u_k)$:]
  $\ell_i = (3i+4n-j-3, 1, 3i+4n-k-1)$
\end{description}
The last sublist consists of a single $7n$.

All $n + 2$ lists described above are concatenated, resulting in
a single list of $7n-2$ numbers.

If there is a path from $u_1$ to $u_n$ in $G'$, we find a way through the
purgatory as follows. We necessarily start on the first number and
jump to the central number (necessarily a 1) in the first sublist of length 3. The index
of the current sublist corresponds to the last vertex $u_s$ in the portion of
the path traversed so far. We inductively extend the portion by moving
one position back or forth depending on the edge $(u_s, u_t)$ chosen in
the path in $G'$. Now the current number by construction takes us to
position $t$ and then to the $t$-th sublist. If eventually vertex $u_n$
is reached, the position in the purgatory is one after the list and the
construction terminates. Conversely, the only transitions possible
in the purgatory are those corresponding to edges since the numbers
traversed are at least $3n$. Therefore the only jumps possible are those
desribed above.
\qed

%\bibliographystyle{alpha}
%\bibliography{kopfnuss} 

\end{document}